\documentclass[conference]{IEEEtran}
\IEEEoverridecommandlockouts
\usepackage{amsmath,epsfig}
\usepackage{textcomp}
\usepackage{algorithm,algorithmic}

\usepackage{amsmath,graphicx}
\usepackage{scrextend}
   \usepackage[square, comma, sort&compress,numbers]{natbib}

\usepackage{color,soul}

\definecolor{gray}{rgb}{0.7,0.7,0.7}

\usepackage[small]{caption}
\usepackage{booktabs}
\usepackage{eqparbox}
\usepackage{multirow}
\usepackage{flushend}

\graphicspath{{figures/}}
\def\BibTeX{{\rm B\kern-.05em{\sc i\kern-.025em b}\kern-.08em
    T\kern-.1667em\lower.7ex\hbox{E}\kern-.125emX}}

\begin{document}

\title{Enhancing VMAF through New Feature Integration and Model Combination}

\author{Fan Zhang\textsuperscript{\dag}, Angeliki Katsenou\textsuperscript{\dag}, Christos Bampis\textsuperscript{\S} , Luk\'{a}\v{s} Krasula\textsuperscript{\S},
Zhi Li\textsuperscript{\S} and  David Bull\textsuperscript{\dag} \\
\textit{\textsuperscript{\dag}Bristol Vision Institute, University of Bristol, Bristol, UK, BS1 5DD.} \\
\{fan.zhang,angeliki.katsenou, dave.bull\}@bristol.ac.uk \\ 
\textit{\textsuperscript{\S}Netflix Inc., Los Gatos, CA, USA, 95032.}\\
\{christosb,lkrasula,zli\}@netflix.com}

\maketitle

\begin{abstract}
VMAF is a machine learning based video quality assessment method, originally designed for streaming applications, which combines multiple quality metrics and video features through SVM regression. It offers higher correlation with subjective opinions compared to many conventional quality assessment methods. In this paper we propose enhancements to VMAF through the integration of new video features and alternative quality metrics (selected from a diverse pool) alongside multiple model combination. The proposed combination approach enables training on multiple databases with varying content and distortion characteristics. Our enhanced VMAF method has been evaluated on eight HD video databases, and consistently outperforms the original VMAF model (0.6.1) and other benchmark quality metrics, exhibiting higher correlation with subjective ground truth data. 
\end{abstract}


\section{Introduction}
\label{sec:intro}

In recent years, the consumption of video data has increased significantly. It was predicted by CISCO that, by 2022, 80\% of the world's data traffic will be video \cite{r:cisco2}. However, since 2020, due to the impact of the COVID-19 pandemic, it is clear that this is an underestimate. Video quality assessment is thus a key tool for video service providers, enabling them to accurately estimate the quality of their encoded content and hence improve the experience of their users. Quality assessment plays a critical role in benchmarking encoding methods, comparing configurations and for creating optimum bit rate ladders prior to adaptive streaming over networks with variable bandwidth to devices with differing capabilities.  

Perceptual video quality can be assessed through psychophysical experiments which, while effective, are costly and time consuming. Objective quality metrics offer more efficient solutions, but to be effective these must be designed to achieve good correlation with subjective results. Objective quality assessment methods can be classified into three primary categories according to how much information they utilise from the reference (original) material, (i) full reference (FR) (ii) reduce reference and (iii) no-reference. In this paper, we only focus on full reference video quality metrics. 

The most commonly used FR metric is PSNR (peak signal-to-noise ratio). PSNR simply measures the pixel-wise distortions between the test content and its reference counterpart, and thus does not always correlate well with visual perception. Over the past two decades, many perceptually inspired objective quality metrics have been proposed, targeting enhanced correlation with subjective quality compared to PSNR. Notable examples include SSIM \cite{j:ssim} and its variants \cite{j:vssim,c:mssim}, VIF \cite{j:Sheikh}, VSNR \cite{j:vsnr}, VQM \cite{j:Pinson}, MOVIE \cite{j:movie}, MAD \cite{j:MAD}, STMAD \cite{c:STMAD} and PVM \cite{j:Zhang3}. Further details on objective quality assessment methods can be found in references \cite{j:Chikkerur1,b:Zhang2}. 

In recent years, improved quality assessment performance has been achieved using machine learning techniques. One of the most successful examples is the Video Multimethod Assessment Fusion (VMAF) approach, developed by Netflix. This combines several existing quality metrics and video features including ADM \cite{j:Li2}, VIF (at four different scales) and the average temporal frame difference (TI), using a support vector machine (SVM) regressor. VMAF was trained on a large HD video quality database, VMAF+ \cite{w:VMAF}, and has been reported to outperform conventional quality assessment methods on various subjective databases. It also includes various extensions developed for different viewing scenarios such as UHDTV and cellular phones. 

However, VMAF only includes one temporal feature (TI) which is based on simple frame differencing and this can lead to inconsistent performance when applied to video content with complex temporal activity (e.g. dynamic textures) \cite{j:Zhang7}. Additionally, the training database VMAF+ is limited as it only contains test sequences generated by an H.264 codec and resolution re-sampling. The trained fusion model, therefore, cannot be guaranteed to achieve optimal performance when used on content compressed by other video codecs.

In this context, inspired by previous work on dynamic texture classification \cite{j:Zhang} and model fusion \cite{j:Bampis}, we propose a VMAF enhancement method based on three primary modifications: (i) the employed ADM metric is enhanced by integrating a feature related to dynamic textures; (ii) two new SVM models are trained based on selection from a pool of original VMAF and new features; (iii) the final quality index is obtained by linearly combining these two models, where training is based on two training databases. The enhanced VMAF method has been benchmarked against the original VMAF and against other popular quality metrics using eight evaluation databases. Results show consistent correlation improvement with subjective ground truth on all test datasets.

The remainder of this paper is organised as follows. Section \ref{sec:algo} describes the proposed algorithm, focusing on the details of three primary modifications. Section \ref{sec:cfg} presents the experimental configurations, including training/test materials and evaluation metrics. Section \ref{sec:results} provides the comparison results between the proposed method and benchmark approaches on the test content alongside the complexity analysis. Finally, the conclusions and future work are outlined in Section \ref{sec:conclusion}.

\section{Proposed Algorithm}
\label{sec:algo}

\begin{figure}[t]
  \centerline{\includegraphics[width=\linewidth]{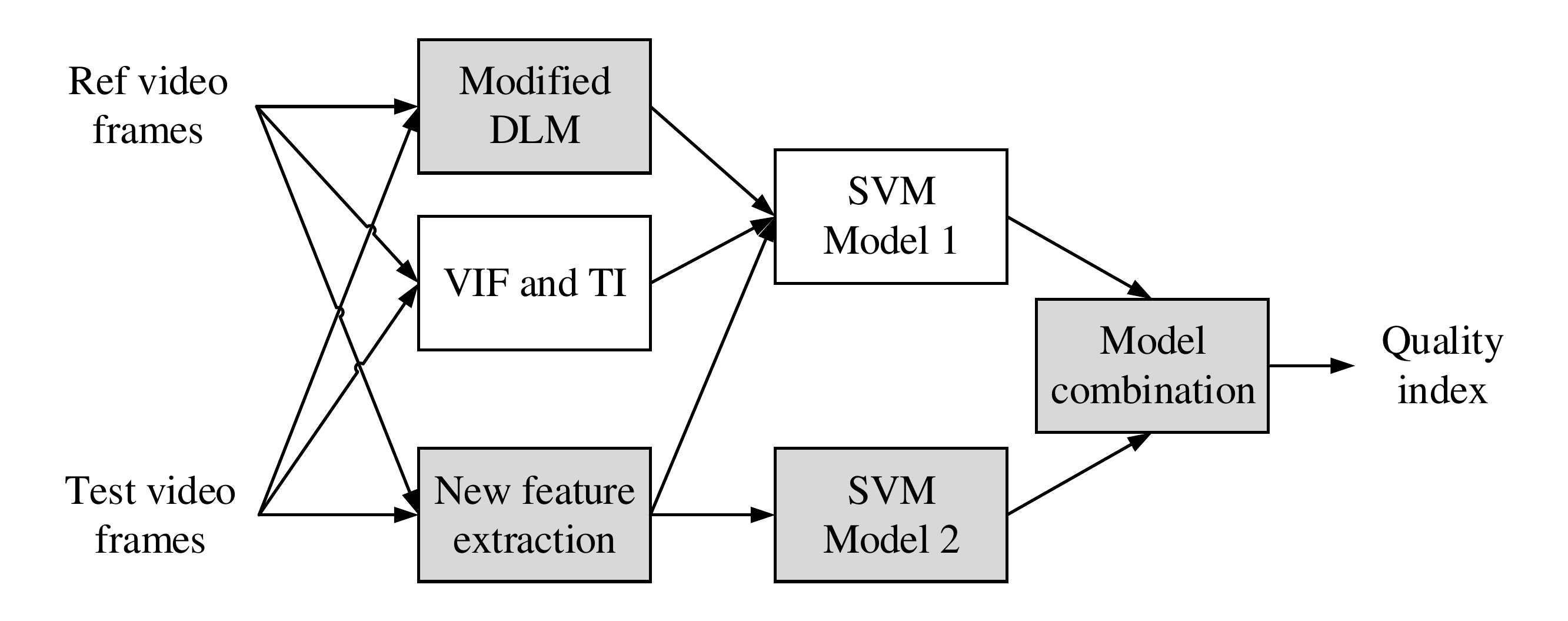}}
\caption{Diagrammatic illustration of the proposed approach.}
\label{fig:workflow}
\end{figure}

The proposed approach is illustrated in Fig. \ref{fig:workflow}. Compared to the original VMAF method, the new approach calculates a modified ADM index based on the input reference and test video frames, which will be fused with temporal frame difference (TI), four VIF values (at four different resolution scales) and new selected video features using a re-trained SVM regression model (Model 1). A second SVM model (Model 2) was trained based on new selected features and a different training dataset to provide a further estimate of the test video quality. The outputs from these two SVM models are combined using a normalised linear model to generate the final quality index. The implementation details relating to each of these components are described below.

\subsection{Dynamic Texture Feature Integration}
\label{sec:DLM}

ADM \cite{j:Li2} is an image quality assessment method which predicts perceptual quality through separately estimating detail losses (DLM) and additive impairments (AIM). The calculation of the detail losses exploit two HVS characteristics -- contrast sensitivity and spatial masking. However the current ADM model is limited in that it does not fully capture temporal masking effects (for example due to dynamic textures, e.g. water, smoke, and steam). Therefore, based on the dynamic texture classification method proposed in \cite{j:Zhang}, we have modified the contrast masking thresholds $\mathbf{MT}_\lambda$ (equation (18) of \cite{j:Li2}) in ADM to include weighting by a dynamic texture feature ($\mathbf{DTF}_\lambda$) as shown below:
\begin{equation}
\mathbf{MT}_{\lambda,\mathrm{new}}= \frac{\mathbf{MT}_\lambda}{(1+\mathbf{DTF}_\lambda)^\alpha}
\end{equation}
Here $\mathbf{MT}_{\lambda,\mathrm{new}}$ are the new masking thresholds which replace the original $\mathbf{MT}_\lambda$ in the ADM calculation. $\lambda$ is the wavelet decomposition level defined in \cite{j:Li2}. $\alpha$ is a parameter which was empirically obtained to achieve the optimal correlation performance (in terms of the Spearman Rank Order Correlation) with subjective ground truth based on the VMAF+ database \cite{w:VMAF}. 

$\mathbf{DTF}_\lambda$ is a dynamic texture feature based on the displaced frame difference (luma component only) after motion compensation (using the previous frame as reference). The motion compensation operation is performed using an optical flow approach (Lucas-Kanade Method) \cite{c:Lucas}.

\begin{equation}
\mathbf{DTF} = \left|\mathbf{o}-\mathbf{o}_\mathrm{DF}\right|
\label{eq:DD1}
\end{equation}
Here we use notation similar to that in \cite{j:Li2} - $\mathbf{o}$ represents the original frame and $\mathbf{o}_\mathrm{DF}$ represents the displaced frame after motion compensation. $\mathbf{DTF}$ is decomposed using a 2D discrete wavelet transform (DWT) at different scales to calculate the corresponding $\mathbf{DTF}_\lambda$ at the scale $\lambda$.

\subsection{Additional Feature Selection}

In order to achieve improved correlation performance, new feature candidates (alongside the original VMAF features) are computed. These include popular quality metrics and video features as summarised in TABLE \ref{tab:feature}. These features are calculated separately on luma and chroma channels (if applicable) and at four different resolution scales (low resolution frames are obtained through 2D DWT decomposition). This results in a total of 165 new features. 

\begin{table}[htbp]
\centering
\small
\caption{Feature candidates for fusion.}
\begin{tabular}{p{2.5cm} | p{5.5cm}}
\toprule
Feature Type & Feature Name\\
\midrule 
\midrule Quality metrics & PSNR, SSIM \cite{j:ssim}, MS-SSIM \cite{c:mssim}, VSNR \cite{j:vsnr}, VIF \cite{j:Sheikh}, MAD \cite{j:Chandler} and PVM \cite{j:Zhang3} \\
\midrule Image and video features & SI \cite{j:Winkler1}, CF \cite{j:Winkler1}, TP \cite{j:Zhang7}, $\Delta$SI, $\Delta$CF, $\Delta$TI, $\Delta$TP, LUMA, BL, ED\\
\bottomrule		
\end{tabular}
\label{tab:feature}
\end{table}

The definitions of most features included in TABLE \ref{tab:feature} can be found in their references. Image and video features SI, CF, TP, LUMA and HIST are calculated based on reference sequences, while $\Delta$SI, $\Delta$CF, $\Delta$TI and $\Delta$TP are the average feature difference between the test sequences and their original counterparts. LUMA, 
 BL and ED are three new features: LUMA captures the average luminance values while BL and ED are features that estimate blurring and edge artefacts in the test content. Their formulae are defined below.
\begin{equation}
						\mathbf{BL}(x,y) = \sum_{H,V,D}{|\mathbf{DWT}_\mathrm{o}(x,y)|} - \sum_{H,V,D}{|\mathbf{DWT}_\mathrm{t}(x,y)|}
\end{equation}
\begin{equation}
						\mathbf{BL}(x,y)= 0, \mathrm{if} \ \mathbf{BL} (x,y) <0
\end{equation}
\begin{equation}
						\mathbf{ED}(x,y) = \sum_{H,V,D}{|\mathbf{DWT}_\mathrm{t}(x,y)|} - \sum_{H,V,D}{|\mathbf{DWT}_\mathrm{o}(x,y)|}
\end{equation}
\begin{equation}
						\mathbf{ED}(x,y) = 0, \mathrm{if} \  \mathbf{ED}(x,y) <0
\end{equation}
Here, the blurring artefact $\mathbf{BL}$ is similar to that in \cite{j:Zhang3}, which calculates the energy loss in three high frequency DWT subbands ($H$, $V$, $D$) of the test video frames. The edge artefact $\mathbf{ED}$ is the additional energy present in the high frequency DWT subbands compared to the reference coefficients.

During feature selection, we followed an approach which is similar to the Sequential Forward Method Selection (SFMS) method in \cite{c:Lin}, and used the Spearman Rank Order Correlation Coefficient (SROCC) as the performance measurement. The algorithm description is provided in Algorithm \ref{alg:SFMS}.

\begin{algorithm}[ht]
\footnotesize
\caption{The employed feature selection method}
\label{alg:SFMS}
\begin{algorithmic}[1]
\REQUIRE \ \\
Candidate feature pool: $\mathbf{F}=\{f_1, f_2, \ldots , f_N\}$;\\
Existing selected feature set: $\mathbf{F^*}$;\\
Training database: $D_T$.
\ENSURE \ \\
Optimal feature set: $\mathbf{F^{**}}$ by maximising SROCC\\
\hrulefill
\STATE Train a SVM regression model $s$ based on $\mathbf{F^*}$ and $D_T$;
\STATE Initialise $J= \mathrm{SROCC}(\mathbf{F^*}, D_T)$ (a SVM model is trained on $D_T$ to combine all the features in $\mathbf{F^*}$);
\STATE Initialise the output feature set $\mathbf{F^{**}}=\mathbf{F^*}$
\WHILE{$\mathbf{F}\neq \emptyset$} 
		\STATE Pick the next best feature: \\
		$f_i=\mathrm{argmax}_{f_i\in (\mathbf{F}-\mathbf{F^{**}})} \mathrm{SROCC}(\mathbf{F^{**}}+f_i, D_T)$
		\IF{$\mathrm{SROCC}(\mathbf{F^{**}}+f_i, D_T)>J$}
				\STATE Update \\
				$J= \mathrm{SROCC}(\mathbf{F^{**}+f_i}, D_T)$;\\
				$\mathbf{F^{**}}=\mathbf{F^{**}}+f_i$ and $\mathbf{F}=\mathbf{F}-f_i$;
		\ELSE
				\STATE Break the while loop;	
					\ENDIF
\ENDWHILE 
\RETURN $\mathbf{F^{**}}$
\end{algorithmic}  
\end{algorithm}

For SVM Model 1 in Figure \ref{fig:workflow}, the selected feature set includes the original six VMAF features. The ADM metric was replaced by the enhanced version as described in Section \ref{sec:DLM}. For SVM Model 2, the input selected feature set is empty.

\subsection{Model Combination}

Frame level outputs from the two SVM models are linearly combined to generate the final quality index for each frame. 

\begin{equation}
						Q = \beta\cdot M_1 + (1-\beta)\cdot M_2.
\end{equation}

The weighting parameter $\beta \in[0,1]$ was selected to achieve the best overall SROCC value on training datasets.

\section{Experimental Configuration}
\label{sec:cfg}

Ten video quality databases including HD (1920$\times$1080) content have been used for training and evaluating the proposed approach. As we solely focus on distortions and artefacts introduced through compression, non-compression distortion versions have been removed from the datasets. However, because resolution re-sampling is commonly used in many video streaming applications, test sequences with re-sampling artefacts have been retained. The databases employed here, their compression codecs, and distortions types are summarised in TABLE \ref{tab:database}. Among these databases, VMAF+ is used to train the SVM regression model 1 ($M_1$) (as it was the training dataset used for the original VMAF model). CC-HDDO was selected to train the second SVM model ($M_2$), because it contains video content compressed by both HEVC and AV1 codecs as well as resolution re-sampling artefacts. The other eight databases are used for evaluation.

\begin{table}[htbp]
\centering
\caption{Ten HD subjective video databases employed.}
\begin{tabular}{r || c |c |c |c}
\toprule
Database  & SRC & Size & Codec& Artefacts\\
\midrule 
\midrule  VMAF+  \cite{w:VMAF} & 23 & 230 &  x264&C+R \\
\midrule  CC-HDDO \cite{c:Zhang24}& 9 & 90 & HM/AV1 & C+R\\
\midrule 
\midrule NFLX-P \cite{w:VMAF}& 9 & 70& x264 & C+R\\
\midrule  NFLX \cite{w:VMAF}& 33& 259 &x264 & C+R\\
\midrule  MCL-V \cite{j:Lin4}& 12&96& x264 & C+R\\
\midrule   BVI-HD \cite{j:Zhang7}& 32&192&HM &C\\
\midrule   CC-HD \cite{j:Zhang16}& 9&108&HM/AV1/VTM & C\\
\midrule   SHVC \cite{r:JCTVCW0095}& 9&64&HM& C\\
\midrule   IVP \cite{w:IVP}& 10 & 100&Dirac/JM/MPEG-2 & C\\
\midrule   VQEGHD3 \cite{r:vqegHD}& 13& 72& MPEG-2/JM&C\\
\bottomrule		
\end{tabular}
\flushleft
Note: C: Coding; R: Re-sampling
\label{tab:database}
\end{table}

\subsection{Evaluation and Benchmarking}

The subjective scores associated with the eight test databases were used to evaluate the performance of the proposed method. For reference and further comparison, eight other popular objective quality metrics have also been tested on these databases. These are: PSNR, SSIM \cite{j:ssim}, MS-SSIM \cite{c:mssim}, VIF \cite{j:Sheikh}, VSNR \cite{j:vsnr}, ADM \cite{j:Li2} VMAF (0.6.1) \cite{w:VMAF} and ST-VMAF \cite{j:Bampis}. 

The performance of all metrics was evaluated using the Spearman Rank Order Correlation Coefficient (SROCC). A significance test was also conducted to identify the difference in performance between the original VMAF (0.6.1) and other tested metrics on all test databases. The approach in \cite{j:movie,j:Zhang3} was used whereby an F-test was conducted on the residual between the average MOS (mean opinion scores) of each database and the MOS predicted by the tested objective quality metrics through a non-linear regression using a logistic function.

\subsection{Ablation Study}

Our three primary contributions have also been tested on the eight test datasets and compared to the full version.

(1) \textbf{ADM modification (denoted as w/o E-ADM)}: the effectiveness of the enhanced ADM method (E-ADM) has been evaluated by replacing it with the original ADM \cite{j:Li2}. Other  selected features remain the same, and the SVM models have been re-trained on the same training databases.

(2) \textbf{New feature selection (denoted as w/o NF)}: new selected features are substituted by original VMAF features (except E-ADM), and two models are trained separately on two databases based on the same  (original VMAF) features. The enhanced ADM metric is used here.

(3) \textbf{Model combination (denoted as $M_1$ and $M_2$)}: the correlation performance of two separate SVM models are presented and compared to the combined version.

\section{Results and Discussions}
\label{sec:results}

The new selected feature sets for the two SVM models are listed in TABLE \ref{tab:featureset}, where $M_1$ corresponds to the extension of the original VMAF and $M_2$ is the second model based on new selected features. The weighting parameter $\beta$ is 0.5. Table \ref{tab:results} presents a summary of performance comparison between the proposed approach and the other eight benchmark quality metrics. For each evaluated quality metric, the SROCC value on each test dataset is presented alongside an aggregate SROCC (Overall) value for all eight databases. We followed the method in \cite{j:Bampis} to calculate the aggregate correlation coefficient using Fisher transformation \cite{j:Corey}, in which the SROCC value for each database was transformed based on:
\begin{equation}
z_\mathrm{SROCC}=\frac{1}{2} \mathrm{ln} \frac{1+\mathrm{SROCC}}{1-\mathrm{SROCC}} 
\label{eq:ztrans}
\end{equation}
The transformed values are then averaged and inverse transformed to obtain the aggregate SROCC.

\begin{table}[htbp]
\centering
\footnotesize
\caption{The two new feature sets selected using the method in Algorithm \ref{alg:SFMS}. Here feature notation A-C-Si stands for feature A applied on channel C (YCbCr space) at scale i (i=1, $\ldots$, 4). Original VMAF features are highlighted in bold font.}
\begin{tabular}{l|| p{7cm} }
\toprule
Model   & Selected features \\ 
 \midrule $M_1$& E-ADM, \textbf{TI-Y-S3}, \textbf{VIF-Y-S1}, \textbf{VIF-Y-S2}, \textbf{VIF-Y-S3}, \textbf{VIF-Y-S4}, ED-Y-S4, BL-Y-S2\\ 
 \midrule $M_2$& E-ADM, \textbf{TI-Y-S3}, VIF-Cb-S1, $\Delta$TI-Cb-S4, PSNR-Y-S4, $\Delta$SI-Cr-S1 \\ 
\bottomrule		
\end{tabular}
\label{tab:featureset}
\end{table}

\begin{table*}[htbp]
\centering
\footnotesize
\caption{Performance of the proposed method (including ablation study variants) and other benchmark approaches on eight test databases. The values in each cell $x (y)$ correspond to the SROCC value ($x$) and F-test result ($y$) at 95\% confidence interval. $y$=1 suggests that the metric is superior to VMAF 0.6.1 ($y$=-1 if the opposite is true), while $y$=0 indicates that there is no significant difference between them.}
\begin{tabular}{r || llllllll | c  }
\toprule
SROCC (F-test)   & NFLX & NFLX-P & BVI-HD & CC-HD & MCL-V & SHVC & IVP & VQEGHD3& Overall  \\ 
	 \midrule
 \midrule PSNR & 0.6218 (-1) & 0.6596 (-1) & 0.6143 (-1) & 0.6166 (-1) & 0.4640 (-1) & 0.7380 (-1) & 0.8168 (-1) & 0.7518 (-1) & 0.6729  \\ 
\midrule SSIM& 0.5638 (-1) & 0.6054 (-1) & 0.5992 (-1) & 0.7194 (-1) & 0.4018 (-1) & 0.5446 (-1) & 0.7139 (-1) & 0.7361 (-1) & 0.6209  \\ 
\midrule MS-SSIM & 0.7136 (-1) & 0.7394 (-1) & 0.7652 (0) & 0.7534 (-1) & 0.6306 (0) & 0.8007 (-1) & 0.8330 (0) & 0.8457 (-1) & 0.7675  \\ 
\midrule VSNR & 0.7873 (-1) & 0.8612 (-1) & 0.7408 (0) & 0.5763 (-1) & 0.5988 (-1) & 0.6661 (-1) & 0.8124 (-1) & 0.7753 (-1) & 0.7417  \\ 
\midrule VIF & 0.7784 (-1) & 0.8818 (-1) & 0.7712 (0) & 0.7459 (-1) & 0.7380 (0) & 0.7367 (-1) & 0.7489 (-1) & 0.8623 (0) & 0.7901  \\ 
\midrule ADM & 0.9272 (0) & 0.9252 (0) & 0.7699 (0) & 0.6858 (-1) & 0.6994 (0) & 0.9107 (0) & 0.8836 (0) & 0.8565 (0) & 0.8549  \\ 
\midrule VMAF 0.6.1 & 0.9254 & 0.9104 & 0.7962 & 0.8723 & 0.7766 & 0.9114 & 0.8786 & 0.8442 & 0.8730\\
\midrule ST-VMAF & 0.9354 (0) & \textbf{0.9270} (0) & 0.7682 (0) & 0.7727 (0) & 0.8079 (0) & 0.9228 (0) & 0.8689 (0) & \textbf{0.8925} (0) & 0.8619\\
\midrule \midrule Proposed  & \textbf{0.9433} (1) & 0.9253 (0) & \textbf{0.8057} (0) & \textbf{0.8783} (0) & \textbf{0.8282} (0) & \textbf{0.9241} (0) & \textbf{0.9022}(0) & 0.8796 (0) & \textbf{0.8940}\\ \midrule
\midrule w/o E-ADM& 0.9400 (0) & 0.9173 (0) & 0.8016 (0) & 0.8826 (0) & 0.8145 (0) & 0.9228 (0) & 0.9012 (0) & 0.8779 (0) & 0.8905\\
\midrule w/o NF&  0.8655 (-1) & 0.8917 (0) & 0.7997 (0) & 0.8866 (0) & 0.8122 (0) & 0.8489 (-1) & 0.8159 (0) & 0.8358 (0) & 0.8478\\
\midrule $M_1$& 0.9338 (0) & 0.9168 (0) & 0.8067 (0) & 0.8595 (0) & 0.8044 (0) & 0.9221 (0) & 0.9060 (0) & 0.8652 (0) & 0.8853 \\
\midrule $M_2$  & 0.9418 (1) & 0.9141 (0) & 0.7920 (0) & 0.8376 (0) & 0.8327 (0) & 0.8729 (0) & 0.8810 (0) & 0.8591 (0) & 0.8743\\
\bottomrule		
\end{tabular}
\label{tab:results}
\end{table*}

It can be observed that the proposed method offers higher aggregate SROCC values across the eight test databases compared to the other quality metrics. Furthermore, improvement has been achieved over the original VMAF 0.6.1 for all eight test databases, and this result is statistically significant (based on the F-test) on the NFLX dataset. Moreover, according to the ablation study results in TABLE \ref{tab:results}, all three primary contributions have led to higher overall correlation performance when compared to their tested replacements. The newly selected features contribute the largest improvement.  

\begin{figure}[htbp]
\footnotesize
\centering
\begin{minipage}[b]{0.485\linewidth}
\centering
\centerline{\includegraphics[width=1.1\linewidth]{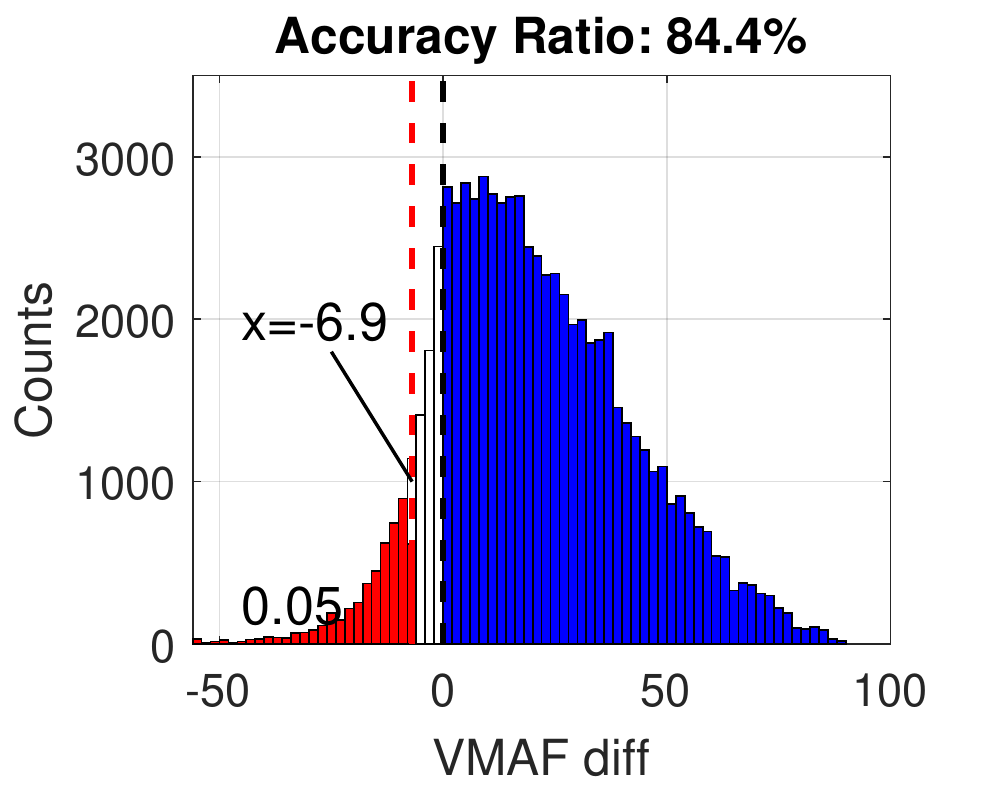}}
\end{minipage}
\begin{minipage}[b]{0.485\linewidth}
\centering
\centerline{\includegraphics[width=1.1\linewidth]{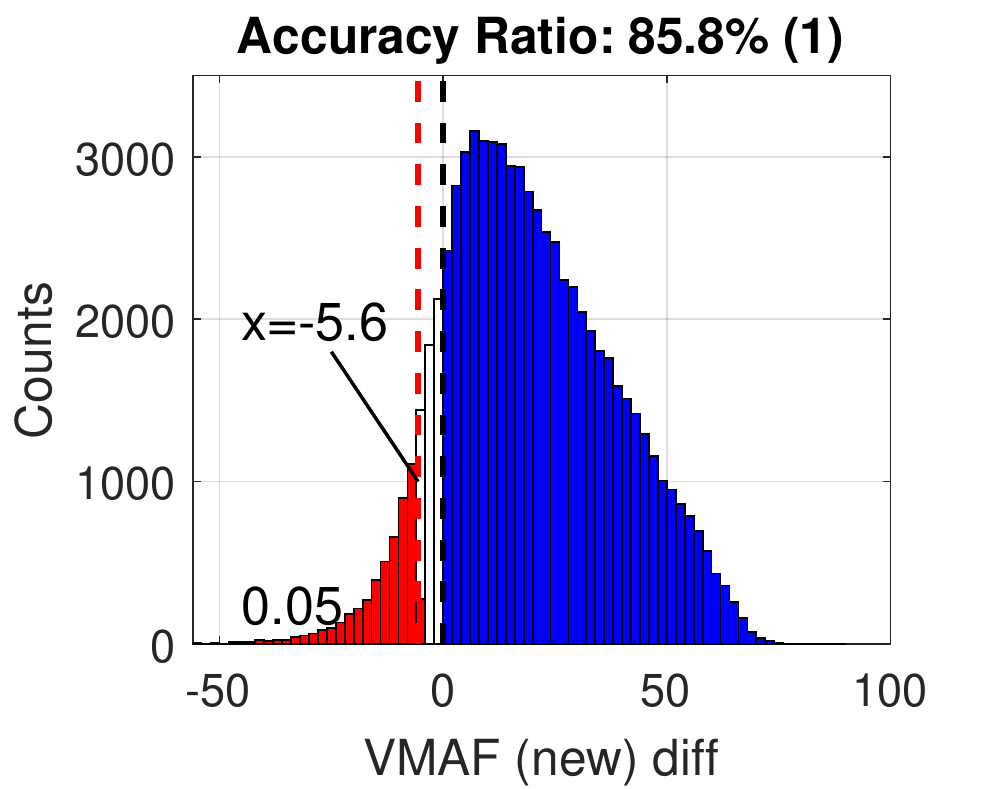}}
\end{minipage}
\caption{The distribution of quality metric difference values for the original VMAF (left) and the proposed method (right). The 5\% percentiles are indicated in both figures.}
\label{fig:cmb}
\end{figure}

When quality metrics are employed in practical applications, it is useful to compare performance using distorted video sequences generated by different codecs and coding configurations. Inspired by the database combination idea proposed in \cite{j:Krasula}, we have calculated subjective MOS (after the ranges have been normalised to 0-100) difference between every pair of test sequences in each of the eight test databases. When computing the MOS difference, we artificially changed the order to ensure that MOS differences are equal to, or greater than, zero. This results in MOS difference scores corresponding to 102,842 test sequence pairs. The quality index difference between the test sequences in every pair has also been computed (following the same order). The distributions of the quality index difference values are shown in Fig. \ref{fig:cmb} for both the original VMAF and the proposed method. We also calculate the accuracy ratio for both metrics, which is defined as the number of pairs where the VMAF difference values provide the same classification results as the MOS differences. It can be observed that the proposed approach offers the higher accuracy ratio (85.8\%) compared to the original VMAF (84.4\%) - the improvement (1.4\%) represents an additional 1440 (102842$\times$1.4\%) video pairs that have been correctly classified. This improvement is significant according to Fisher's exact test ($p=1.2\times 10^{-11}$). 

\section{Conclusion}
\label{sec:conclusion}

In this paper, we have proposed an enhanced version of VMAF based on dynamic texture feature integration (into ADM), additional feature selection and model combination. Multiple training databases have also been used to train the SVM fusion models. The new approach shows consistent improvements over the original VMAF model on all test databases in terms of correlation with subjective ground truth. It also achieves superior performance when  used to compare different distorted versions of the same sources on a large combined dataset. Future work will focus on the more challenging cases (e.g. the outliers in Fig. \ref{fig:cmb}) when multiple codecs and various coding configurations are employed.

\small
\bibliographystyle{IEEEtran}
\bibliography{IEEEabrv,MyRef}

\begin{thebibliography}{10}
\providecommand{\url}[1]{#1}
\csname url@samestyle\endcsname
\providecommand{\newblock}{\relax}
\providecommand{\bibinfo}[2]{#2}
\providecommand{\BIBentrySTDinterwordspacing}{\spaceskip=0pt\relax}
\providecommand{\BIBentryALTinterwordstretchfactor}{4}
\providecommand{\BIBentryALTinterwordspacing}{\spaceskip=\fontdimen2\font plus
\BIBentryALTinterwordstretchfactor\fontdimen3\font minus
  \fontdimen4\font\relax}
\providecommand{\BIBforeignlanguage}[2]{{%
\expandafter\ifx\csname l@#1\endcsname\relax
\typeout{** WARNING: IEEEtran.bst: No hyphenation pattern has been}%
\typeout{** loaded for the language `#1'. Using the pattern for}%
\typeout{** the default language instead.}%
\else
\language=\csname l@#1\endcsname
\fi
#2}}
\providecommand{\BIBdecl}{\relax}
\BIBdecl

\bibitem{r:cisco2}
CISCO, ``{CISCO} visual networking index: forecast and methodology,
  2017--2022,'' November 2018.

\bibitem{j:ssim}
Z.~Wang, A.~Bovik, H.~Sheikh, and E.~Simoncelli, ``Image quality assessment:
  from error visibility to structural similarity,'' \emph{IEEE Transactions on
  Image Processing}, vol.~13, pp. 600--612, 2004.

\bibitem{j:vssim}
Z.~Wang, L.~Lu, and A.~C. Bovik, ``Video quality assessment based on structural
  distortion measurement,'' \emph{Signal Processing: Image Communication},
  vol.~19, no.~2, pp. 121--132, 2004.

\bibitem{c:mssim}
Z.~Wang, E.~P. Simoncelli, and A.~C. Bovik, ``Multi-scale structural similarity
  for image quality assessment,'' in \emph{Proc. Asilomar Conference on
  Signals, Systems and Computers}, vol.~2.\hskip 1em plus 0.5em minus
  0.4em\relax IEEE, 2003, p. 1398.

\bibitem{j:Sheikh}
H.~R. Sheikh, A.~C. Bovik, and G.~de~Veciana, ``An information fidelity
  criterion for image quality assessment using natural scene statistics,''
  \emph{IEEE Transactions on Image Processing}, vol.~14, pp. 2117--2128, 2005.

\bibitem{j:vsnr}
D.~Chandler and S.~Hemami, ``{VSNR}: A wavelet-based visual signal-to-noise
  ratio for natural images,'' \emph{IEEE Transactions on Image Processing},
  vol.~16, no.~9, pp. 2284--2298, 2007.

\bibitem{j:Pinson}
M.~H. Pinson and S.~Wolf, ``A new standardized method for objectively measuring
  video quality,'' \emph{IEEE Transactions on Broadcasting}, vol.~50, pp.
  312--322, 2004.

\bibitem{j:movie}
K.~Seshadrinathan and A.~C. Bovik, ``Motion tuned spatio-temporal quailty
  assessment of natural videos,'' \emph{IEEE Transactions on Image Processing},
  vol.~19, pp. 335--350, 2010.

\bibitem{j:MAD}
E.~C. Larson and D.~M. Chandler, ``Most apparent distortion: full-reference
  image quality assessment and the role of strategy,'' \emph{Journal of
  Electronic Imaging}, vol.~19, no.~1, pp. 011\,006(1--21), 2010.

\bibitem{c:STMAD}
P.~V. Vu, C.~T. Vu, and D.~M. Chandler, ``A spatiotemporal
  most-apparent-distortion model for video quality assessment,'' in \emph{Proc.
  IEEE Int Conf. on Image Processing}.\hskip 1em plus 0.5em minus 0.4em\relax
  IEEE, 2011, pp. 2505--2508.

\bibitem{j:Zhang3}
F.~Zhang and D.~R. Bull, ``A perception-based hybrid model for video quality
  assessment,'' \emph{IEEE Transactions on Circuits and Systems for Video
  Technology}, vol.~26, no.~6, pp. 1017--1028, 2016.

\bibitem{j:Chikkerur1}
S.~Chikkerur, V.~Sundaram, M.~Reisslein, and L.~J. Karam, ``Objective video
  quality assessment methods: A classification, review, and performance
  comparison,'' \emph{IEEE transactions on broadcasting}, vol.~57, no.~2, pp.
  165--182, 2011.

\bibitem{b:Zhang2}
D.~Bull and F.~Zhang, \emph{Intelligent Image and Video Compression:
  Communicating Pictures}, 2nd~ed.\hskip 1em plus 0.5em minus 0.4em\relax
  Academic Press, In Press.

\bibitem{j:Li2}
S.~Li, F.~Zhang, L.~Ma, and K.~H. Ngan, ``Image quality assessment by
  separately evaluating detail losses and additive impairments,'' \emph{IEEE
  Transactions on Multimedia}, vol.~13, no.~5, pp. 935--949, 2011.

\bibitem{w:VMAF}
Z.~Li, A.~Aaron, I.~Katsavounidis, A.~Moorthy, and M.~Manohara, ``Toward a
  practical perceptual video quality metric,'' \emph{The Netflix Tech Blog},
  2016.

\bibitem{j:Zhang7}
F.~Zhang, F.~{Mercer Moss}, R.~Baddeley, and D.~R. Bull, ``{BVI-HD}: A video
  quality database for {HEVC} compressed and texture synthesised content,''
  \emph{IEEE Transactions on Multimedia}, vol.~20, no.~10, pp. 2620--2630,
  October 2018.

\bibitem{j:Zhang}
F.~Zhang and D.~R. Bull, ``A parametric framework for video compression using
  region-based texture models,'' \emph{IEEE Journal of Selected Topics in
  Signal Processing}, vol.~5, no.~7, pp. 1378--1392, 2011.

\bibitem{j:Bampis}
C.~G. Bampis, Z.~Li, and A.~C. Bovik, ``Spatiotemporal feature integration and
  model fusion for full reference video quality assessment,'' \emph{IEEE
  Transactions on Circuits and Systems for Video Technology}, vol.~29, no.~8,
  pp. 2256--2270, 2018.

\bibitem{c:Lucas}
B.~Lucas and T.~Kanade, ``An iterative image registration technique with an
  application to stereo vision,'' in \emph{Proceedings of the International
  Joint Conference on Aritifcial Intelligence}, 1981.

\bibitem{j:Chandler}
D.~M. Chandler, ``Seven chanllenges in image quality assessment: Past present,
  and future research,'' \emph{ISRN Signal Processing}, vol. 2013, 2013.

\bibitem{j:Winkler1}
S.~Winkler, ``Analysis of public image and video database for quality
  assessment,'' \emph{IEEE Journal of Selected Topics in Signal Processing},
  vol.~6, no.~6, pp. 1--10, 2012.

\bibitem{c:Lin}
J.~Y. Lin, T.-J. Liu, E.~C.-H. Wu, and C.-C.~J. Kuo, ``A fusion-based video
  quality assessment (fvqa) index,'' in \emph{Signal and Information Processing
  Association Annual Summit and Conference (APSIPA), 2014 Asia-Pacific}.\hskip
  1em plus 0.5em minus 0.4em\relax IEEE, 2014, pp. 1--5.

\bibitem{c:Zhang24}
A.~V. Katsenou, F.~Zhang, M.~Afonso, and D.~R. Bull, ``A subjective comparison
  of {AV1} and {HEVC} for adaptive video streaming,'' in \emph{Proc. IEEE Int
  Conf. on Image Processing}, 2019, pp. 4145--4149.

\bibitem{j:Lin4}
J.~Y. Lin, R.~Song, C.-H. Wu, T.~Liu, H.~Wang, and C.-C.~J. Kuo, ``{MCL-V}: A
  streaming video quality assessment database,'' \emph{Journal of Visual
  Communication and Image Representation}, vol.~30, pp. 1--9, 2015.

\bibitem{j:Zhang16}
F.~Zhang, A.~V. Katsenou, M.~Afonso, G.~Dimitrov, and D.~R. Bull, ``Comparing
  {VVC}, {HEVC} and {AV1} using objective and subjective assessments,''
  \emph{arXiv preprint arXiv:2003.10282}, 2020.

\bibitem{r:JCTVCW0095}
Y.~He, Y.~Ye, F.~Hendry, Y.-K. Wang, and V.~Baroncini, ``{SHVC} verification
  test results,'' in \emph{the JCT-VC meeting}, no. JCTVC-W0095.\hskip 1em plus
  0.5em minus 0.4em\relax {ITU-T, ISO/IEC}, 2016.

\bibitem{w:IVP}
\BIBentryALTinterwordspacing
F.~Zhang, S.~Li, L.~Ma, Y.~C. Wong, and K.~N. Ngan, ``{IVP} subjective quality
  video database,'' 2009. [Online]. Available:
  \url{http://ivp.ee.cuhk.edu.hk/research/database/subjective/}
\BIBentrySTDinterwordspacing

\bibitem{r:vqegHD}
\BIBentryALTinterwordspacing
{Video Quality Experts Group}, ``Report on the validation of video quality
  models for high definition video content,'' 2010. [Online]. Available:
  \url{http://www.its.bldrdoc.gov/vqeg/projects/hdtv/hdtv.aspx}
\BIBentrySTDinterwordspacing

\bibitem{j:Corey}
D.~M. Corey, W.~P. Dunlap, and M.~J. Burke, ``Averaging correlations: Expected
  values and bias in combined pearson rs and fisher's z transformations,''
  \emph{The Journal of general psychology}, vol. 125, no.~3, pp. 245--261,
  1998.

\bibitem{j:Krasula}
L.~Krasula, Y.~Baveye, and P.~Le~Callet, ``Training objective image and video
  quality estimators using multiple databases,'' \emph{IEEE Transactions on
  Multimedia}, vol.~22, no.~4, pp. 961--969, 2019.

\end{thebibliography}

\end{document}